# Flexoelectric and piezoelectric coupling in a bended MoS$_2$ monolayer

Hanna V. Shevliakova[1,2], Semen O. Yesylevskyy[1,3], Ihor Kupchak[4], Galina I. Dovbeshko[1], Yunseok Kim[5] and Anna N. Morozovska[1*]

[1] Institute of Physics of the National Academy of Sciences of Ukraine, pr. Nauky 46, 03028 Kyiv, Ukraine.

[2] Department of Microelectronics, Ihor Sikorsky Kyiv Polytechnic Institute, Kyiv, Ukraine

[3] Laboratoire Chrono Environnement UMR CNRS 6249, Université de Bourgogne Franche-Comté, Besançon, France.

[4] Institute of Semiconductor Physics of NAS of Ukraine, Kyiv, Ukraine

[5] School of Advanced Materials Science and Engineering, Sungkyunkwan University (SKKU), Suwon 16419, Republic of Korea

[*]Corresponding author: anna.n.morozovska@gmail.com

**Abstract**

Low-dimensional (LD) transition metal dichalcogenides (TMDs) in the form of nanoflakes, which consist of one or several layers, are the subject of intensive fundamental and applied research. Due to the size-induced transition from a bulk to nanoscale, they can be both nonpolar, piezoelectric or even ferroelectric. Also, in terms of electronic properties, they can be direct-band semiconductors, semi-metals or even metals. The tuning of the electronic properties in the LD-TMDs are commonly related with applied strains and strain gradients, which can affect strongly their polar properties via the piezoelectric and flexoelectric couplings. Using the density functional theory (DFT) and phenomenological Landau approach, we studied the bended 2H-MoS$_2$ monolayer and analyzed its flexoelectric and piezoelectric properties. The dependences of the dipole moment, strain and strain gradient on the coordinate along the layer were calculated. From these dependences the components of the flexoelectric and piezoelectric tensors have been determined and analyzed. Obtained results are useful for applications of LD-TMDs in strain engineering and flexible electronics.

# I. INTRODUCTION

Layered transition metal dichalcogenides (**TMDs**) in the form of bulk materials are typically non-polar centrosymmetric semiconductors with a relatively wide band gap ~(1.1 – 2) eV and specific form of Fermi surface [1, 2]. On transition from the bulk to the nanoscale additional long-range orderings and physical properties, such as piezoelectric, or even ferroelectric [3, 4, 5], semiconductive, semi-metallic or metallic [6, 7, 8] were found in different structural phases (polymorphs) of TMD monolayers [9]. In particular, the polar and semiconducting properties of the low-dimensional (**LD**) TMDs with a chemical formula $MX_2$ (M – metal Mo, V, W; X – chalcogen S, Se, Te) [10, 11] and Janus-compounds (**JC**) with a chemical formula MXY (X, Y – chalcogens) [12, 13] are varying from non-polar to ferroelectric state, and from direct-band semiconductor to metallic conductivity.

The strain and strain gradient impact on LD-TMD polar and electronic properties can be principally important for the properties control, and therefore LD semiconductor materials, such as graphene, $MX_2$ and MXY monolayers, are ideal candidates for the strain engineering [14] and recently introduced "straintronics" [15]. Their strain-induced conductive domain walls can act as mobile charged channels, similarly to the "domain wall nanoelectronics" in multiferroic thin films [16, 17, 18] and graphene-on-ferroelectric nanostructures [19, 20]. There are remarkable possibilities for tuning the structural, polar and electronic properties of LD-TMDs by application of either homogeneous elastic strains [21, 22] or inhomogeneous curvature-induced strain gradients [23, 24]. In particular, Duerloo et. al. [6] predicted a strain-induced phase transition from a semiconducting 2H to a metallic 1T' phase in various $MX_2$. Later on, and Song et. al. [25] observed a room temperature semiconductor−metal transition in thin $MoTe_2$ films induced by a homogeneous tensile strain of 0.2%.

A number of first-principles studies explored the surface-induced piezoelectricity [26, 27] and ferroelectric polarization [5, 28] in various $MX_2$ and MXY. Later on, the possible mechanism of the ferroelectric state appearance in LD-TMDs was described by the Landau-Ginzburg-Devonshire (**LGD**) continuous approach [29]. Since only in-plane ferroelectricity can exist in a geometrically flat pure centrosymmetric $MX_2$ layer, LGD analysis suggests that the switchable out-of-plane ferroelectric polarization emerges due to the rare-earth doping, and predicts that the domain walls in LD-TDMs should become conductive above a certain strain threshold.

The physical origin of the bending-induced changes of LD-TMDs in electronic and polar properties can be similar to the ones in a bended graphene [30, 31] and boron nitride [32], at that the flexoelectric effect plays a very important role. Indeed, it was predicted theoretically, that the bending can induce an out-of-plane electric dipole moment of carbon nanoshells [33]. As anticipated, the bending-induced dipole moment is curvature-dependent, and strong curvatures can lead to a relatively high polarization of LD-TMDs, in fact induced by a flexoelectric effect [34, 35]. Actually, the bending-induced out-of-plane dipole moment with density $p \sim (0.01 - 0.4)$C/nm and flexoelectric polarization $P \sim (0.1 - 2)\mu$C/cm$^2$ were calculated from the first principles for MoS$_2$ [5], WS$_2$ [35], and WTe$_2$ [36, 37] single-layers. In contrast to dielectrics and wide gap ferroelectrics, where the flexoelectricity is determined by the lattice deformation, in centrosymmetric narrow gap semiconductors and semimetals, such as MX$_2$, contains a significant contribution from the deformed electronic density [32, 38, 39] that, in principle, can dominate over the lattice-mediated ionic contribution especially under photoexcitation [40, 41].

The tunable out-of-plane piezoelectricity and enhanced conductivity, both induced by a flexoelectricity, were observed by Kang et al. [23] in semiconducting 2H-MoTe$_2$ flakes by creating the surface corrugation. The experimental results were corroborated by their *ab initio* calculations [23], analytical calculations performed within LGD approach [42] and finite element modeling (**FEM**) [43]. Specifically, LGD approach explores the flexoelectric origin of the polarization induced by a spontaneous bending and by inversion symmetry breaking due to the interactions with substrate. FEM allows calculating the elastic and electric fields, flexoelectric polarization, and its correlation with free charge density for a TMD (or JC) nanoflake placed on a rough substrate with a sinusoidal profile of the corrugation [43].

Despite the progress, the complete information about the piezoelectric and flexoelectric coupling tensors in LD-TMDs is lacking, and the influence of the tensor symmetry and numerical values of its components on polar and electronic phenomena in LD-TMDs has not been studied. Since the couplings are This knowledge is required to control and predict the physical properties of LD-TMDs and JC for their novel applications in nanoelectronics and advanced memories. Using the density functional theory (**DFT**) and phenomenological Landau approach, in this work we consider the curved monolayer of 2H-MoS$_2$ in order to calculate the nonzero components of its flexoelectric and piezoelectric coupling tensors and to analyze their possible dependence on the layer corrugation and surface-induced symmetry lowering.

## II. THEORETICAL FORMALISM

The static electric polarization of a LD-TMD, $P_i(\vec{x})$, has the form:

$$P_i(\vec{x}) \cong f_{ijkl} \frac{\partial u_{kl}}{\partial x_j} + \frac{e^s_{ijk}}{t} u_{jk} - \varepsilon_0 \chi_{ij} \frac{\partial \varphi}{\partial x_j}. \qquad (1)$$

Here $u_{ij}(\vec{x})$ is the elastic strain tensor, $f_{ijkl}$ is the static flexoelectric tensor [44] determined by the microscopic properties of the material [45, 46], $e^S_{ijk}$ is the tensor of the surface-induced piezoelectric effect [47, 48], and $t$ is the layer or nanoflake thickness. The last term in Eq.(1) is proportional to the electric field, $E_i = -\frac{\partial \varphi}{\partial x_i}$, where $\varphi(\vec{x})$ is an electric potential, $\varepsilon_0$ is a universal dielectric constant, $\chi_{ij}$ is a real part of the TMD dielectric susceptibility. Einstein summation over repeated indexes is used hereinafter.

### A. Ab initio calculations of the flat and corrugated MoS$_2$ nanolayer

We performed the calculations of the atomic position in the MoS$_2$ nanolayer for the case of a flat layer and for (1 – 10)% corrugated layers with a 1% step in corrugation within the DFT [49] in the generalized gradient approximation (GGA), implemented in the Quantum Espresso code [50]. We have used ultrasoft Perdew-Burke-Ernzerhof (PBE) pseudopotentials [51], which include 14 valence electrons for molybdenum and 6 valence electrons for sulfur. An integration of the Brillouin zone has been performed using 14×1×1 $\boldsymbol{k}$-points mesh centered on $\Gamma$ in the Brillouin zone, generated by Monkhorst-Pack scheme [52], and Methfessel-Paxton smearing [53] with a parameter of 0.005 Ry. We applied 50 Ry cutoff for smooth part of the wave function and 350 Ry for the augmented charge density to ensure a sufficient convergence of the results.

The MoS$_2$ monolayer was modeled by a supercell approach with 25 Å of vacuum layer added to avoid a coulomb interaction between the periodic images. Initial (flat) layer was built with experimental value of the lattice constant of 3.161 Å and contained 48 atoms (**Fig. 1**). For the corrugated layers, we reduced the lattice constant of the supercell in $x$-direction and modulated the atomic $z$ (out-of-plane) coordinates by suitable sinusoidal distribution to match an integer value of corrugation. After that, all the systems were relaxed through all the internal coordinates until the Hellmann-Feynman forces became less than $10^{-4}$ atomic units, and at this point all the necessary quantities (atomic coordinates and wave functions) were extracted. It is worth to note, that the difference between the initial corrugation and that obtained for the corresponding relaxed system does not exceed few percent of their values. Therefore, the calculations revealed that the bending in $z$-direction is sinusoidal along the $x$-axis (**Fig. 2**).

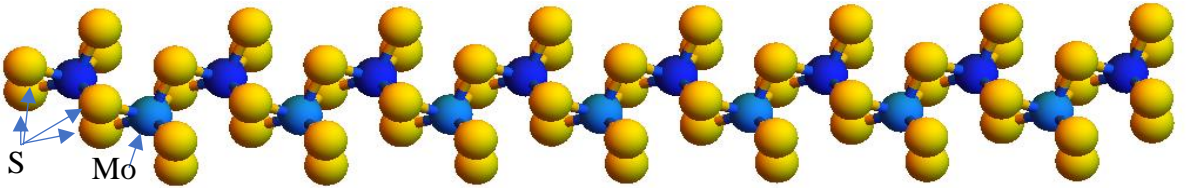

**FIGURE 1.** MoS$_2$ monolayer, where yellow spheres are sulfur (S) atoms and blue spheres are molybdenum (Mo) atoms.

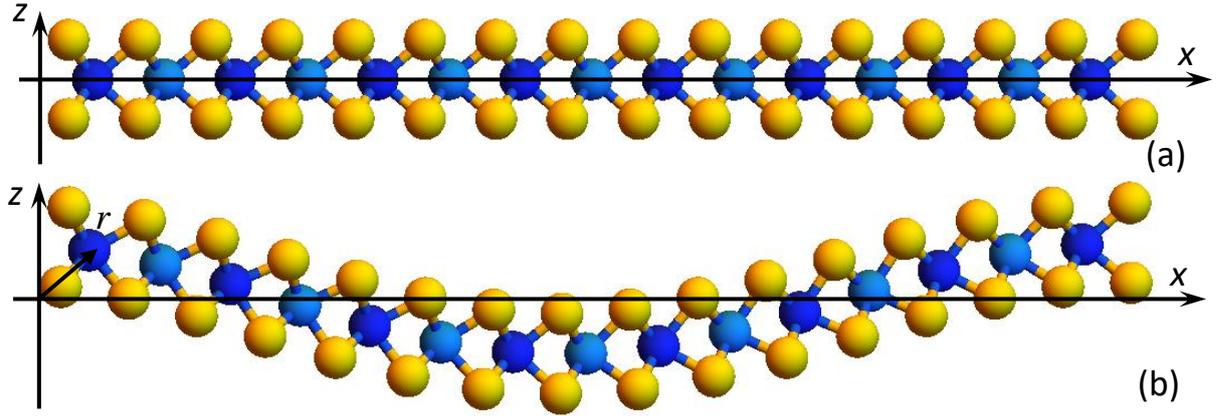

**FIGURE 2.** Side view of the MoS$_2$ monolayer (a) without bending and (b) with 10% bending.

In addition to the position of the atoms, the Bader charges [54] were computed. The distribution of the charges appeared to be uniform in a flat system and equal to $Q_{Mo} = 1,083e$ for molybdenum atoms, and is $Q_S = $ -0,5415 $e$ for sulfur atoms (in the units of elementary charge $e$). For strongly curved layers with a bending (8–10)%, a harmonic dependence of the charges on the $x$ coordinate was observed (as shown in **Fig. 3**). However, the difference between the maximal and minimal charges appeared to be very small (see the scale in **Fig. 3**) and close to the magnitude of expected numerical error. The effect of such distribution on the electrophysical parameters of the layer is considered neglected hereinafter. This, in fact, means that we neglect the terms proportional to the deformation potential in Eq.(1), as the first approximation, and extract the terms proportional to the net flexoelectric coefficient, $f_{ijkl}$, from the DFT results. The charge variations can be found as the second approximation, which only becomes important for large curvatures [32].

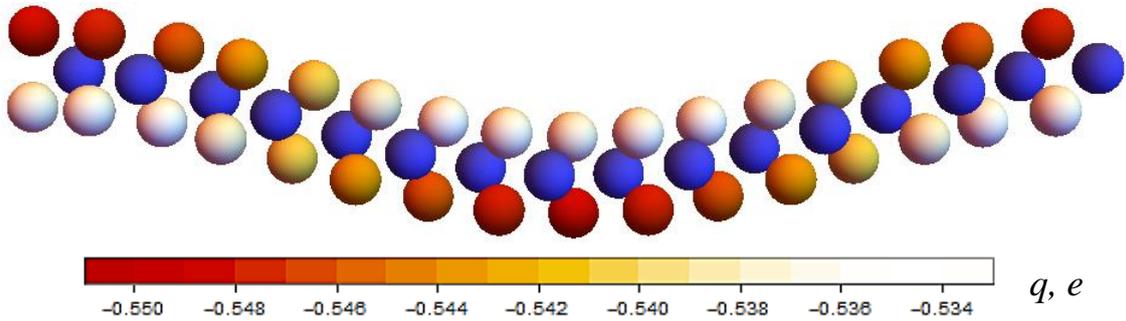

**FIGURE 3**. Charge distribution near sulfur (S) atoms along the chain for a 10% corrugation.

## B. Determination of the flexoelectric coefficients from ab initio calculations

The components of the electrostriction and flexoelectric tensors for the 2H-MoS$_2$, that belongs to the point symmetry group 6/*mmm*, can be determined from the equations:

$$d_z(x) = e_{zzx}u_{zx}(x) + e_{zxx}u_{xx}(x) + f_{zxzx}u_{zx,x}(x) + f_{zxxx}u_{xx,x}(x), \quad (2)$$

$$d_x(x) = e_{xzx}u_{zx}(x) + e_{xxx}u_{xx}(x) + f_{xxxz}u_{xz,x}(x) + f_{xxxx}u_{xx,x}(x), \quad (3)$$

where $d_i$ is projection of the elementary dipole moment on the axis $i$; $u_{ij}$ and $u_{ij,k}$ are strain tensor and strain gradient caused by the layer bending; $e_{ijk}$ and $f_{ijkl}$ are the components of the piezoelectric and flexoelectric tensors.

All nonzero components, $d_i$, $u_{ij}$ and $u_{ij,k}$, were calculated based on the DFT results. In particular, the effective elementary dipole moment $\vec{d}$ of each dipole (see **Fig. 4**) was calculated from the following equation:

$$\vec{d}_i = \vec{r}_{Mo_i} \cdot |Q_{Mo}| - \sum_{j=1}^{6} \vec{r}_{S_{ij}} \cdot \left|\frac{Q_S}{3}\right|, \quad (4)$$

where $\vec{r}_{M_i} = \{x_X, y_X, z_X\}_i$ is a radius-vector of the corresponding *X* atoms.

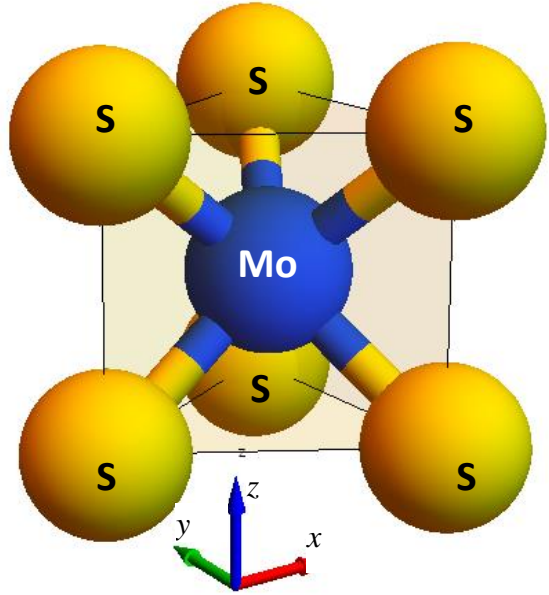

**FIGURE 4**. The atomic system for which the *i*-th effective dipole moment was calculated.

Further dependences were plotted on the *x* coordinate of the mass center of the dipole, and its radius vector was determined by the equation:

$$\vec{r}^{(j,i)} = \frac{1}{m_{Mo} + 6\,m_S}\left(\vec{r}_{Mo}^{(j,i)} m_{Mo} + \sum_{i=1}^{6} \vec{r}_S^{(j,i)} m_S\right), \quad (5)$$

where "*j*" is the percentage of corrugation, "*i*" is the dipole number, $m_{Mo} = 95.9 \frac{g}{mol}$ and $m_S = 32 \frac{g}{mol}$ are atomic masses of molybdenum and sulfur, respectively. The points in **Figs. 5b** and **6b** show the dependence of x-axis and y-axis projection of the dipole moment on x-coordinate of the dipole mass center.

Using the radius vector, the elementary displacement of the mass centers is calculated from the equation:

$$\vec{U}^{(j,i)} = \vec{r}^{(0,i)} - \vec{r}^{(j,i)}, \tag{6}$$

where $\vec{r}^{(0,i)}$ is the radius vector of *i*-th dipole of the undeformed layer, $\vec{r}^{(j,i)}$ is the radius vector of *i*-th dipole of the layer with corrugation *j*%. The dependences of the *x*-axis and *y*-axis projections of the displacement vector $\vec{U}^{(j)}$ on *x* coordinate of the mass center (5) are shown in **Fig. 5a** and **Fig. 6 a**, respectively.

Solid curves in **Fig. 5a-b** and **Fig. 6a-b** are harmonic (sinusoidal and co-sinusoidal) functions, which were selected in the form:

$$d_z = d_{zc}\cos(kx) + d_{zs}\sin(kx), \tag{7a}$$
$$d_x = d_{xs}\sin(2kx) + d_{xc}\cos(2kx) + c. \tag{7b}$$

Here the coefficients $d_{zc}$, $d_{zs}$, $d_{xs}$ and $d_{xc}$ are fitting parameters, which numerical values are selected to reach the best fit of the points in **Fig. 5b** and **Fig. 6b.**

The dependence of the displacement projections on x-axis was approximated by the functions:

$$U_z(x) \approx U_0 \cos(kx), \tag{8a}$$
$$U_x(x) \approx V_0 x + W_0 \cos(2kx). \tag{8b}$$

Here the coefficients $U_0$, $V_0$, and $W_0$ are fitting parameters, which numerical values are selected to reach the best fit of the points in **Fig. 5a** and **Fig. 6a.**

The displacement causes the strain tensor $u_{jk} = \frac{1}{2}\left(\frac{\partial U_k}{\partial j} + \frac{\partial U_j}{\partial k}\right)$, whose components (shown in **Fig. 5c** and **Fig. 6c**) are given by the following equations:

$$u_{zx}(x) = u_{xz}(x) = \frac{1}{2}\frac{\partial U_z}{\partial x} = -\frac{U_0}{2}k\sin(kx), \tag{9a}$$

$$u_{xx}(x) = \frac{\partial U_x}{\partial x} = V_0 + 2W_0 k\cos(2kx). \tag{9b}$$

Their nonzero derivatives (shown in **Fig. 5d** and **Fig. 6d**) are equal to:

$$u_{zx,x}(x) = u_{xz,x}(x) = -\frac{U_0}{2}k^2\cos(kx), \tag{10a}$$

$$u_{xx,x}(x) = \frac{\partial u_{xx}}{\partial x} = -4W_0 k^2 \sin(2kx). \tag{10b}$$

The values $d_{zs}$, $d_{zc}$, $d_{xs}$, $d_{xc}$, $c$, $U_0$, $V_0$, $W_0$ and $k$ in Eqs.(6)-(8) are constants, which are different for each layer. The values of these coefficients are given in **Table I**, and the dependences given by Eqs.(9)-(10) are shown in **Fig. 5c-d** and **Fig. 6c-d**.

Substituting the functions from Eqs.(7) – (10) to Eq.(2)-(3), we obtain:

$$d_{zc}\cos(kx) + d_{zs}\sin(kx) = -e_{331}\frac{U_0}{2}k\sin(kx) + e_{311}[V_0 + 2W_0 k\cos(2kx)] - $$
$$-f_{3131}\frac{U_0}{2}k^2\cos(kx) - f_{3111}4W_0 k^2 \sin(2kx), \qquad (11a)$$

$$d_{xs}\sin(2kx) + d_{xc}\cos(2kx) + c = -e_{131}\frac{U_0}{2}k\sin(kx) + e_{111}[V_0 + 2W_0 k\cos(2kx)] - $$
$$-f_{1113}\frac{U_0}{2}k^2\cos(kx) - f_{1111}4W_0 k^2 \sin(2kx), \qquad (11b)$$

The components of flexoelectric and electrostriction tensors can be obtained from the above expressions:

$$f_{3131} = -\frac{2d_{zc}}{U_0 k^2};\; e_{331} = -\frac{2d_{zs}}{U_0 k};\; e_{311} = 0;\; f_{3111} = 0. \qquad (12a)$$

$$f_{1111} = -\frac{d_{xs}}{4W_0 k^2},\; e_{111} = \frac{d_{xc}}{2W_0 k},\; e_{131} = 0,\; f_{1113} = 0. \qquad (12b)$$

The values of $f_{3131}$, $f_{1111}$, $e_{111}$ and $e_{331}$, calculated according to Eq.(12), taking into account the values of **Table. I**, given in **Table. II.**

Let us underline the dominant contributions of the flexoelectric effect over the piezoelectric effect in both x and z directions, namely $\frac{|f_{ijkl}|}{|e_{ijkl}|} \cong (5-30)\frac{1}{nm}$. This means that a quite realistic strain gradient of about $\frac{1}{nm}$ can induce an order of magnitude higher flexoelectric response in comparison with a piezoelectric reaction. Also note that the dilatational flexoelectric coefficient $f_{1111}$ is almost two times smaller than the shear component $f_{1313}$, while $|e_{331}| > |e_{111}|$.

**Table I.** Fitting parameters

| Corrugation | 1% | 2% | 3% | 4% | 5% | 6% | 7% | 8% | 9% | 10% |
|---|---|---|---|---|---|---|---|---|---|---|
| $k$, 1/nm | 1.44 | 1.441 | 1.443 | 1.446 | 1.449 | 1.453 | 1.457 | 1.463 | 1.468 | 1.475 |
| $d_{xs}$, $10^{-3}$ e·nm | 0.006 | 0.028 | 0.059 | 0.107 | 0.191 | 0.301 | 0.432 | 0.553 | 0.678 | 0.826 |
| $d_{xc}$, $10^{-3}$ e·nm | 0.0001 | -0.006 | -0.004 | -0.059 | -0.062 | -0.048 | -0.039 | -0.078 | -0.142 | -0.166 |
| $c$, $10^{-3}$ e·nm | -0.997 | -0.979 | -0.979 | -0.982 | -0.978 | -1.018 | -1.058 | -1.043 | -1.002 | -1.017 |
| $d_{zs}$, $10^{-3}$ e·nm | 0.066 | 0.122 | 0.174 | 0.325 | 0.341 | 0.382 | 0.448 | 0.530 | 0.596 | 0.67 |
| $d_{zc}$, $10^{-3}$ e·nm | 0.481 | 0.959 | 1.441 | 1.906 | 2.389 | 2.867 | 3.334 | 3.797 | 4.258 | 4.707 |
| $V_0$, $10^{-3}$ a.u. | -0.238 | -0.97 | -2.193 | -3.905 | -6.1 | -8.775 | -11.93 | -15.55 | -19.62 | -24.16 |
| $W_0$, $10^{-3}$ nm | 0.02 | 0.0258 | 0.0449 | 1.296 | 2.06 | 2.883 | 3.852 | 5.019 | 6.356 | 7.791 |
| $U_0$, $10^{-3}$ nm | 20.71 | 41.35 | 61.94 | 82.76 | 103.3 | 123.7 | 144 | 164.1 | 184.2 | 204 |

**Table II.** The components of electrostriction and flexoelectric coefficients

| Curvature, % | $f_{3131}, e/\text{nm}^2$ | $e_{331}, e/\text{nm}$ | $f_{1111}, e/\text{nm}^2$ | $e_{111}, e/\text{nm}$ |
|---|---|---|---|---|
| 1 | -0.0224 | -0.00443 | -0.0338 | 0.00163 |
| 2 | -0.0223 | -0.00409 | -0.131 | -0.0766 |
| 3 | -0.0223 | -0.00389 | -0.156 | -0.0292 |
| 4 | -0.022 | -0.00544 | -0.00983 | -0.0159 |
| 5 | -0.022 | -0.00456 | -0.011 | -0.0103 |
| 6 | -0.022 | -0.00425 | -0.0124 | -0.00573 |
| 7 | -0.0218 | -0.00427 | -0.0132 | -0.00352 |
| 8 | -0.0216 | -0.00442 | -0.0129 | -0.00529 |
| 9 | -0.0214 | -0.00441 | -0.0124 | -0.00759 |
| 10 | -0.0212 | -0.00445 | -0.0122 | -0.00721 |

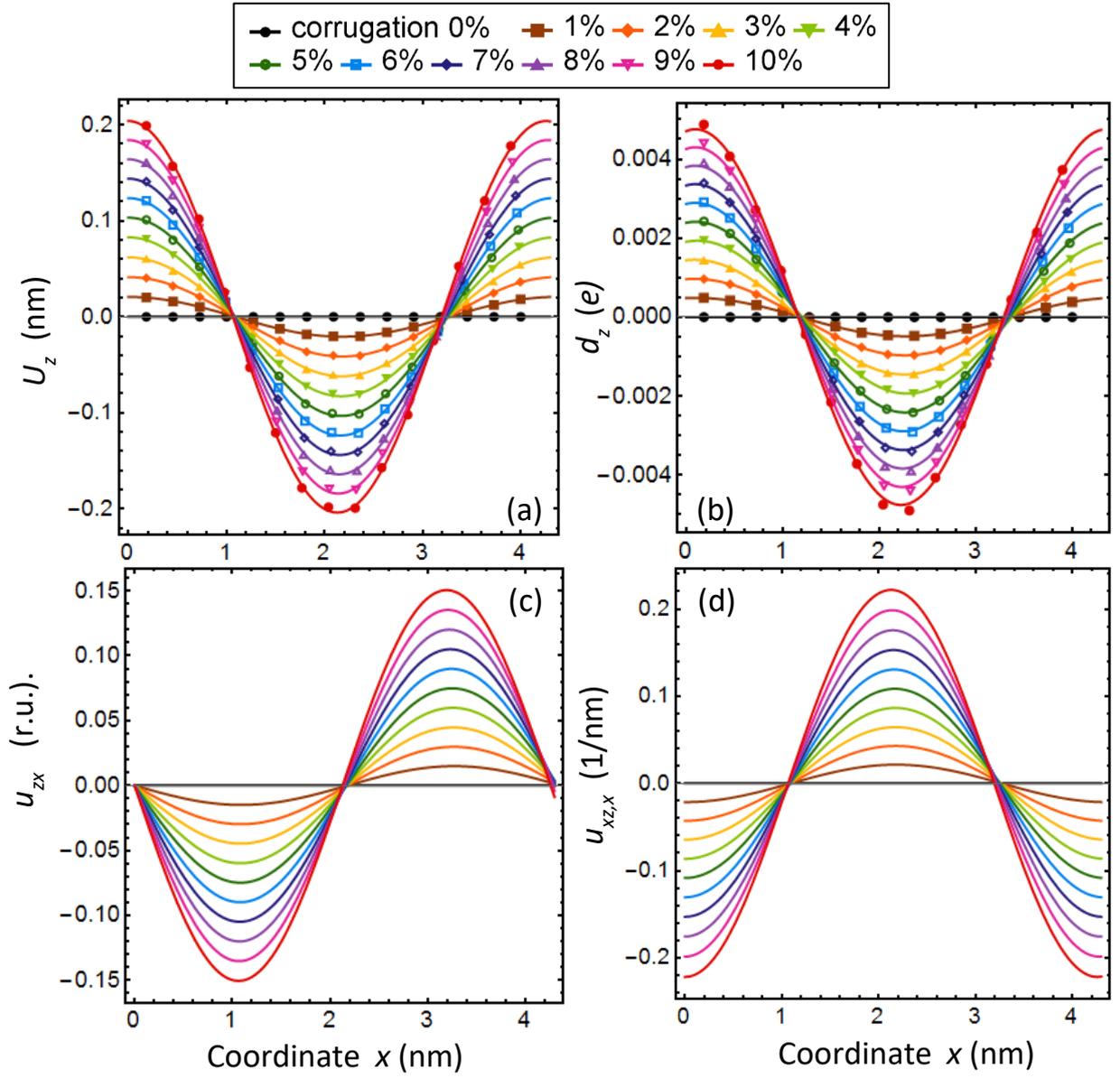

**FIGURE 5.** The dependences on the $x$ coordinate of **(a)** the elementary displacement $U_z$; **(b)** the elementary dipole moment $d_z$; **(c)** the strain tensor component $u_{zx}$ calculated according to Eq.(9a) and **(e)** the strain gradient component $u_{zx,x}$ calculated according to Eq.(10a). The points in the plots **(a)** and **(b)** are discrete values calculated from Eqs.(4)-(6), solid curves are the approximations of these values according to Eqs.(7a) and (8a). All dependences correspond to the effective elementary dipole mass center. The parameters are given in **Table I**.

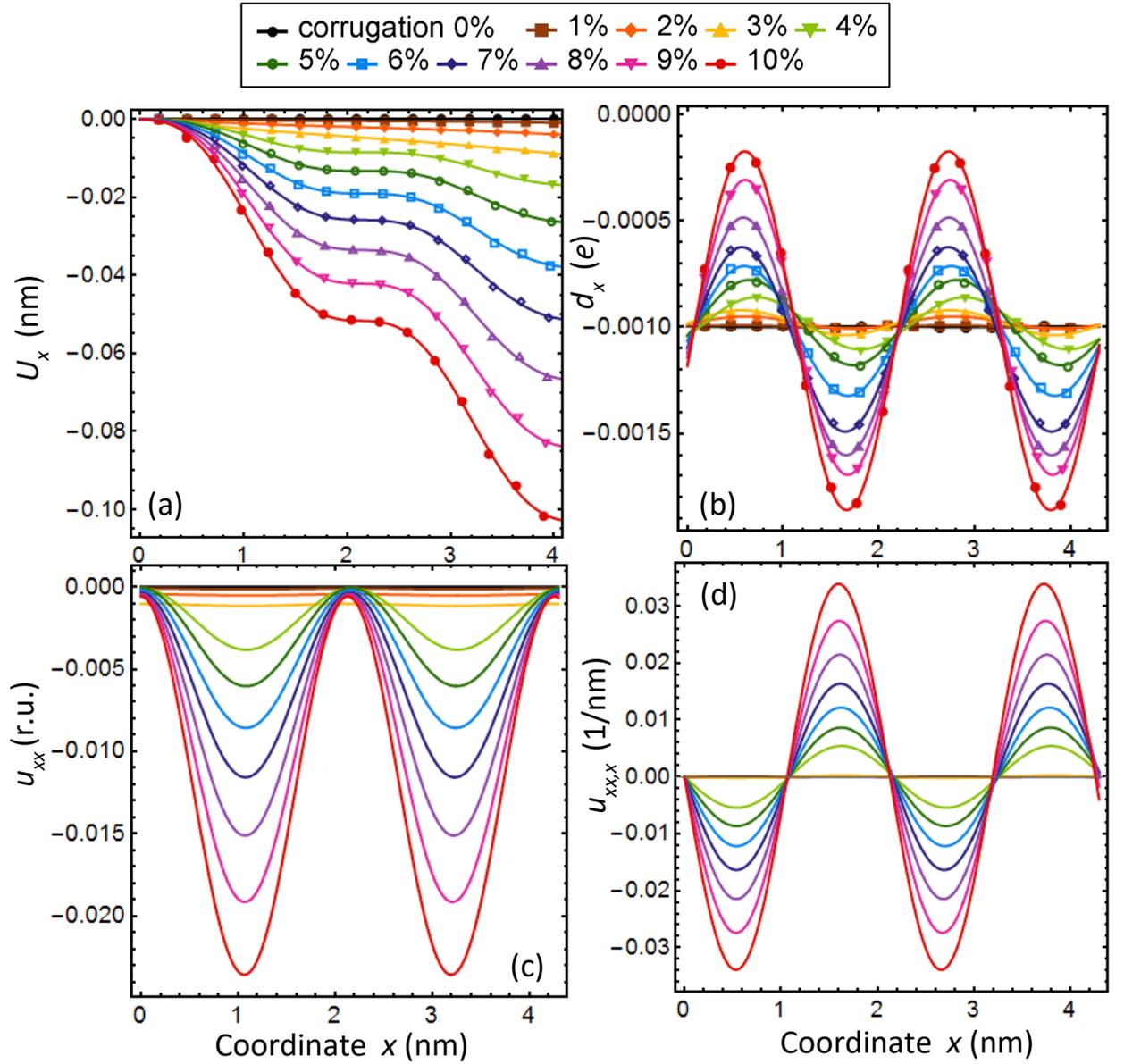

**FIGURE 6.** The dependences on the $x$ coordinate of **(a)** the elementary displacement $U_x$; **(b)** the elementary dipole moment $d_x$; **(c)** the strain tensor component $u_{xx}$ calculated according to Eq.(9b) and **(e)** the strain gradient component $u_{xx,x}$ calculated according to Eq.(10b). The points in the plots **(a)** and **(b)** are discrete values calculated from Eqs.(4)-(6), solid curves are the approximations of these values according to Eqs.(7b) and (8b). All dependences correspond to the effective elementary dipole mass center. The parameters are given in **Table I**.

Dependences of the flexoelectric and piezoelectric tensors components on the corrugation of MoS$_2$ monolayer, which are determined from the DFT calculations, are shown in **Fig. 7**. All components of flexoelectric and piezoelectric tensors can be described by an empirical parabolic dependence, $Ac^2 + Bc + C$ of the corrugation "$c$" (see **Fig. A1** in **Appendix A**). Components $f_{3131}$ and $e_{331}$ weakly depend on corrugation magnitude, and slightly linearly changes with corrugation around a constant value.

Components $f_{1111}$ and $e_{111}$ have an extremum at the corrugation 7%, and reach high absolute values at small corrugations $c < 4\%$. These unreasonably high values originated from the fitting error of the dipole moment projection $d_x(x)$ (see **Fig. A2** in **Appendix A**), so these points should be disregarded.

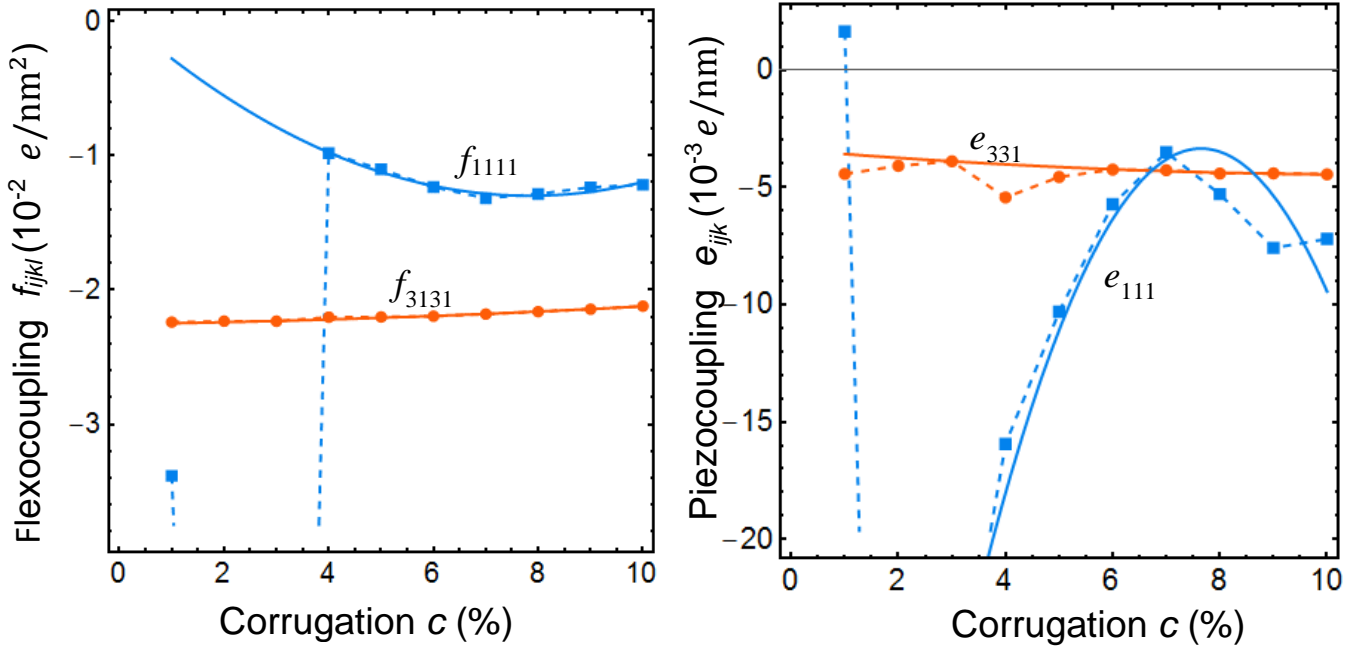

**FIGURE 7.** Dependences of the flexoelectric **(a)** and piezoelectric **(b)** tensors components on the corrugation $c$ of MoS$_2$ monolayer, which are determined from the DFT calculations. Symbols and solid curves correspond to the $f_{1111}$ and $e_{111}$ (blue color), $f_{3131}$ and $e_{331}$ (orange color) dependences determined from DFT calculations and fitted by analytical formulae $Ac^2 + Bc + C$, respectively.

### III. CONCLUSION

Using phenomenological Landau approach combined with DFT *ab initio* calculations, we consider the atomic displacements and charge state of the curved 2H-MoS$_2$ monolayer, and determine the effective components of its flexoelectric and piezoelectric coupling tensors, $f_{ijkl}$ and $e_{ijk}$, as a function of the layer corrugation varying from 0 to 10%.

It appeared that the components of effective flexoelectric and piezoelectric couplings can be described by parabolic dependences of the corrugation. In particular, the components $f_{3131}$ and $e_{331}$ weakly depend on corrugation magnitude, and slightly changes with the corrugation around an almost constant value. The components $f_{1111}$ and $e_{111}$ have an extremum at the corrugation 7%, and reach unreasonably high absolute values at small corrugations < 4%, which are related with the fitting error of the dipole moment projection.

The calculations reveal the dominant contributions of the flexoelectric effect over the piezoelectric effect in both in-plane and out-of-plane directions of the monolayer. This means that a quite realistic strain gradient of about 1 nm$^{-1}$ can induce an order of magnitude higher flexoelectric response in comparison with a piezoelectric reaction. The dilatational flexoelectric coefficient $f_{1111}$ is almost two times smaller than the shear component $f_{1313}$, while $|e_{331}| > |e_{111}|$.

Obtained quantitative results can be useful for elaboration of nanoscale flexible electronic devices based on the bended MX$_2$ layers. In particular, the bended monolayers are promising candidates for the ultra-small diodes and bipolar transistor on MX$_2$, which principal schemes are presented in Ref.[43]. Here the bending profile of the layers controls the sharpness of p-n junctions between the regions with n-type (electron) and p-type (hole) conductivity.

**Authors' contribution.** H.V.S. performed the mathematical treatment and fitting of the DFT results; and prepared figures. S.O.Y. stated the problem for DFT calculations, evolved the procedure of their physical treatment and interpret the results. I.K. performed DFT calculations. A.N.M. and S.O.Y. generated the research idea; A.N.M. evolved the analytical model, interpret the results and wrote the manuscript draft. G.I.D. and Y.K. worked on the manuscript improvement.

**Acknowledgments.** This work (H.V.S., S.O.Y., G.I.D. and A.N.M.) has been supported by the National Research Fund of Ukraine (grant application 2020.02/0027). Y.K. research was supported by Basic Science Research Program through the National Research Foundation of Korea (NRF) funded by the Ministry of Education (No. 2019R1A6A1A03033215).

**APPENDIX A. Additional figures**

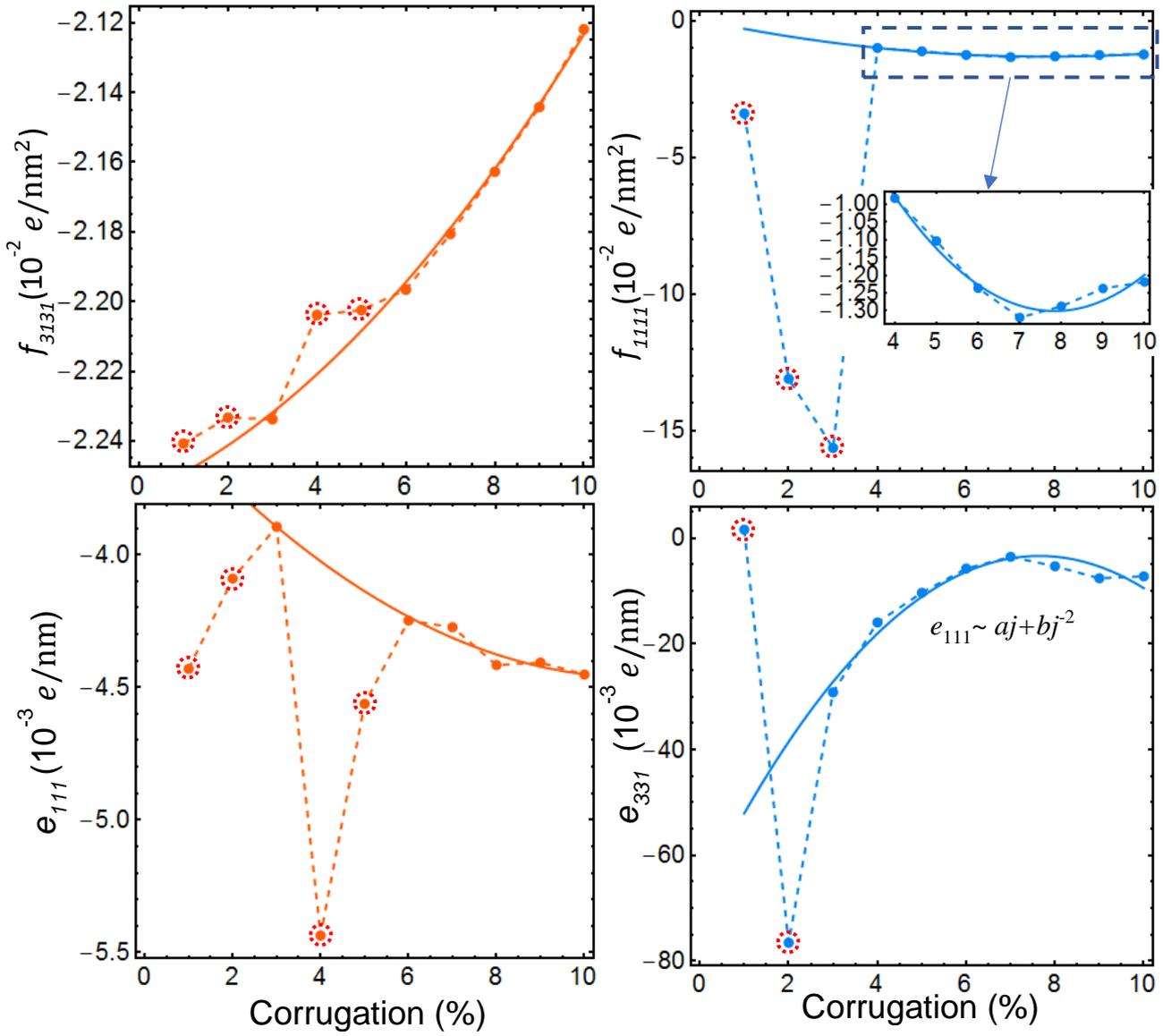

**FIGURE A1.** Dependences of the flexoelectric **(a-b)** and piezoelectric **(c-d)** tensors components on the corrugation of MoS$_2$ layer, which are determined from the DFT calculations. Solid lines present A$x^2$+B$x$+C fitting. Red dotted circles (⊙) indicate points, which were ignored during the fitting.

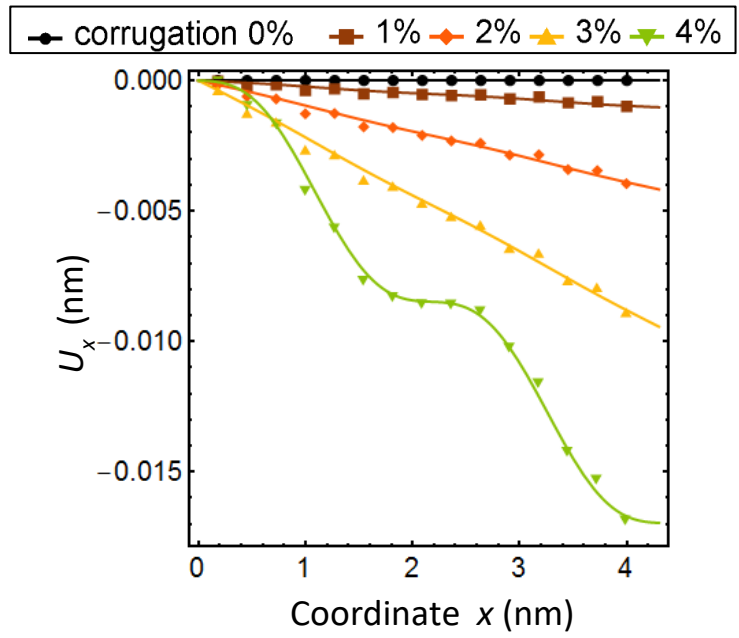

**FIGURE A2.** Dependences of the *x*-axis projections of the elementary displacement on the *x* coordinate of the effective elementary dipole mass center for small corrugations 0-4%.